\newcounter{myctr}

\documentclass{ws-acs}
\usepackage{url}

\renewcommand{\Re}{{\rm Re}\,}

\begin{document}

\makeatletter
\def\@biblabel#1{[#1]}
\makeatother

\markboth{R.~A.~Blythe}{Random copying in space}

%
\catchline{}{}{}{}{}
%

\title{RANDOM COPYING IN SPACE}

\author{\footnotesize RICHARD A.\ BLYTHE}

\address{SUPA, School of Physics and Astronomy, University of Edinburgh\\
Mayfield Road, Edinburgh EH9 3JZ, UK\\
R.A.Blythe@ed.ac.uk}

\maketitle

\begin{history}
\received{(received date)}
\revised{(revised date)}
\end{history}

\begin{abstract}
Random copying is a simple model for population dynamics in the absence of selection, and has been applied to both biological and cultural evolution. In this work, we investigate the effect that spatial structure has on the dynamics. We focus in particular on how a measure of the diversity in the population changes over time.  We show that even when the vast majority of a population's history may be well-described by a spatially-unstructured model, spatial structure may nevertheless affect the expected level of diversity seen at a local scale.  We demonstrate this phenomenon explicitly by examining the random copying process on small-world networks, and use our results to comment on the use of simple random-copying models in an empirical context.
\end{abstract}


\section{Introduction}

Evolution is a theory of change by replication (see e.g.,~\cite{hul88}).  This applies both to biological and cultural evolution, through replication of DNA in the former case, and of practices, behaviors and beliefs in the latter.  Three processes may contribute to the evolutionary dynamics.  Perhaps the most prominent is \emph{selection}, the process by which some individuals in a population may be replicated more often than others.  With no other evolutionary forces acting, the outcome of selection is for the fittest species to outcompete the rest \cite{cro70}.  Greater diversity can be afforded through the introduction of a \emph{mutation} process, which allows the introduction of new, potentially fitter, species into the population.

The third evolutionary process is stochasticity in replication itself, referred to as \emph{drift} by population geneticists \cite{cro70}, and sometimes as \emph{random copying} in a cultural evolution context (see e.g.,~\cite{mes09} for a brief run-down of some recent applications). It is now well understood that, in concert with mutation, a wide range of patterns of diversity can be established through random copying \cite{cro70,hub01}.  In particular, large differences in species abundances can be found, even though they are identical in terms of their birth-death dynamics.  That is, the prevalence of a particular species in a habitat does not necessarily imply that is it any better adapted to that habitat than its competitors. 

On various occasions, good agreement between empirical data and the predictions of these \emph{neutral models}---so called because they lack selection---has been found.  For example, species abundance patterns in tropical forests are well described by a neutral model \cite{con02}, as are various aspects of the dynamics and distribution of baby names in the United States \cite{hah03}. It is sometimes felt that, despite these correspondences between models and data, neutral models lack so much realism that they cannot provide an adequate description of the evolutionary process in question \cite{abr01}.  For example, one might be concerned that neutral models almost always impose a `zero-sum' restriction. That is, every death is assumed to be immediately followed by a birth, so that the population size remains fixed over time.  It is also typically assumed that each individual dies and reproduces at the same rate.  However, the key feature of a neutral model is that it lacks selection, and this does not itself mandate assumptions of the type just outlined.  For example, one can construct neutral models in in which birth and death are independent events, or in which an individual's birth and death rate that may vary with some factor that is uncorrelated with its species.

When applying a neutral theory to empirical data, we are thus drawn to two basic questions.  First, does a good fit to the predictions of a simple neutral model imply that all of its highly restrictive assumptions must be satisfied?  Conversely, does a departure from the predictions of a neutral theory imply that selection must be operating?  In this work, we will argue that the answer to both questions is `no'.  This we achieve by exploiting the unifying theme of this Special Issue: namely, the introduction of \emph{spatial structure} into the random copying dynamics.  This provides one means by which we can relax the assumption that each individual has the same birth and death dynamics as every other.

In a previous work \cite{bly10}, we showed that there are circumstances under which a stochastic equation of motion for the frequency of a species has the same mathematical form even in the presence of nontrivial spatial structure.  Despite this, there are subtle aspects of the dynamics that may differ between the structured and unstructured cases.  Here our aim is to expand on these findings for the less mathematically-inclined reader. We focus on a measure of the expected amount of diversity in the population as a function of time. This quantity, which was mentioned only in passing in \cite{bly10}, turns out to illustrate the subtle effects of spatial structure in a fairly transparent way.  Furthermore, in keeping with the theme of the Special Issue, we mostly have cultural evolutionary applications in mind. In particular, we include some new results for random copying on small-world networks, which can be viewed as a cartoon of cultural evolution by replication across a network of human interpersonal relationships.

The potentially limited role that spatial structure has to play in neutral evolution has long been recognized in population genetics. A prominent idea, dating back to the early work of Wright \cite{wri31}, is that of an \emph{effective population size}. In the current context, this can be thought of as a mapping from a spatially structured model onto one that lacks structure through an appropriate choice of the size of the latter.  There has been considerable work on understanding how different aspects of the spatial structure affect the effective size, and whether different definitions of the effective size are equivalent (see e.g.~\cite{rou04,wan99}).  In particular, it is well understood that different measures of effective size become equivalent when one looks over sufficiently long timescales \cite{ewe04,rou04,whi97}. What seems to have attracted less attention is what counts as ``sufficiently long''. This is what was established in Ref.~\cite{bly10} and discussed in more concrete terms here.  We remark that the formal requirement that there is only one relevant dynamical timescale (illustrated in more detail below) has frequently been assumed elsewhere, for example, in understanding a surprising lack of genetic diversity in spatially-structured habitats \cite{mat06} or in various treatments of consensus times in the socially-inspired \emph{voter model} when put on spatially-structured networks (see e.g.~\cite{soo05,pug09} and in particular the review of \cite{cas09}).

We begin in Section~\ref{unsec} by defining a spatially-unstructured random-copying model, and setting out some of its basic properties.  We then show in Section~\ref{stsec} how to generalize this model to include spatial structure, and explain the main finding of Ref.~\cite{bly10} alluded to above.  In Section~\ref{exsec} we examine explicit examples so as to understand how spatial structure may manifest itself even if a relation to an unstructured model is established.    We conclude in Section~\ref{concs} with a brief summary and some conjectures about the interplay of innovation (mutation) and replication by random copying in a spatial setting.

\section{Spatially Unstructured Random Copying: A Moran Model}
\label{unsec}

\subsection{Model definition}

The two most common formulations of a neutral random-copying process are the \emph{Wright-Fisher} model \cite{fis30,wri31}, in which the entire population is replaced once per timestep, and the \emph{Moran} model \cite{mor58}, in which a single individual is replaced per timestep.  We shall adopt a variant of the Moran model here, whereby instead of a replacement taking place on each tick of a clock, events instead occur as a \emph{continuous-time} (Poisson) process such that, on average, any individual gives birth once per unit time.  After a large number of clock ticks, the difference between the discrete- and continuous-time versions of this process can be disregarded.

For concreteness, we use the example of baby names \cite{hah03} to define the model dynamics.  The system comprises a pool of $M$ names that could be given to a baby of a given gender in some culture.  Any one name can appear multiple times in the pool.  Suppose we are interested in the fate of a particular name, say \emph{Adam}, present at time $t=0$ (i.e., in the beginning).  If there are $m(t)$ instances of this name at some time $t$,  we define the frequency of that name as $x(t) = \frac{m(t)}{M}$.

\begin{figure}[th]
\centerline{\psfig{file=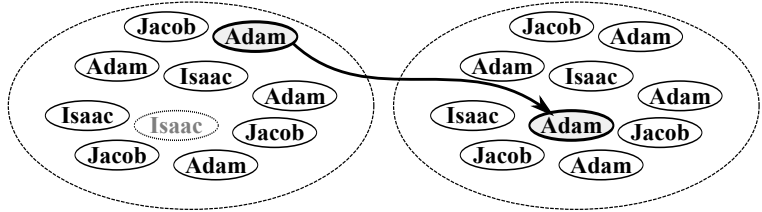,width=10cm}}
\vspace*{8pt}
\caption{\label{fig:babynames} Spatially unstructured random copying dynamics within the `baby names' interpretation \cite{hah03}.  To the left is the pool of names before a sampling event.  An instance of the name `Adam' is selected for copying, displacing a randomly-chosen instance of a name (here, `Isaac') in the process. This gives rise to an updated state of the pool, shown to the right.}
\end{figure}

This frequency may change as a consequence of the following random copying dynamics.  Each instance of a name is \emph{sampled} as a Poisson process at unit rate.  More precisely, this means that in any infinitesimal time interval ${\rm d}t$, any one instance of a name is chosen with probability ${\rm d} t$.  After a sampling event, an existing instance of a name is removed from the pool, and a new copy of the sampled name is placed into the pool. These dynamics are illustrated in Fig.~\ref{fig:babynames}.  In this way, the pool of names serves as some fixed-size `collective memory' of the set of suitable names for children and their relative frequencies.  At any given time, each of the $M$ names in the pool has an equal chance of being the next one to be sampled, and thus to replace some name in the pool. In the above example, the probability that the next child to be named Adam is $x(t)$.

Changes in the frequency of a name happen purely by chance.  For example, $m(t)$ can increase by one if the name sampled is Adam, and the name replaced is not. Equally, it can decrease by one if the name sampled is not Adam, and the name replaced is.  The key point is that the probability of these two events is the same: it is $x(t)[1-x(t)]$ (if we allow for the fact that the instance replaced can be the same as the instance copied).  Therefore the mean change (averaged over multiple realizations of the dynamics) in the frequency of any name is zero.  Changes do nevertheless occur: however these are purely due to random fluctuations.  We remark in passing that the support for this model as an explanation for the dynamics of baby-name frequencies is provided mostly through correspondence between empirical and theoretical distributions \cite{hah03}.  As has been recognized in ecological applications of the same model, stronger support could in principle be obtained through the application of appropriate \emph{sampling formul\ae} (see e.g.,~\cite{alo06}). We return to this point in the conclusion.

\subsection{Decay of diversity in the absence of innovation}

Some versions of a random-copying process include an \emph{innovation} (mutation) step.  In the baby name example, this would correspond to there being some rate at which a completely new name is invented and introduced to the pool, again replacing an instance of an existing name.  In most of this work, we will examine the innovation-free case, although we will return to the topic of innovation in the conclusion.  It almost goes without saying that the model can be applied to other evolutionary examples by a simple relabeling exercise.  More generally, we can think of instances of a name as some kind of \emph{individual} within a population, and the different names as different \emph{species}.  We will use this more general terminology henceforth.

If no innovation is permitted, there are two possible ultimate fates for a species.  Either it can take over the whole population (go to \emph{fixation} in genetics parlance \cite{cro70}), or it goes extinct.  Another way to say the same thing is that the diversity decreases over time.  One way to measure diversity quantitatively is in terms of the probability that, if two individuals are chosen at random, they are of different species. 

One way to determine the expected behavior of this quantity---and that will be of great importance in the discussion of the spatially-structured random-copying process---is to consider the history of this pair of individuals.  Specifically, we can ask the question: how long ago was one of these individuals created as the result of a copying event?  Since each individual is copied at unit rate, and one individual is always replaced whenever this happens, it follows that this creation process (looking backwards in time) is also a Poisson process with unit rate.

\begin{figure}[th]
\centerline{\psfig{file=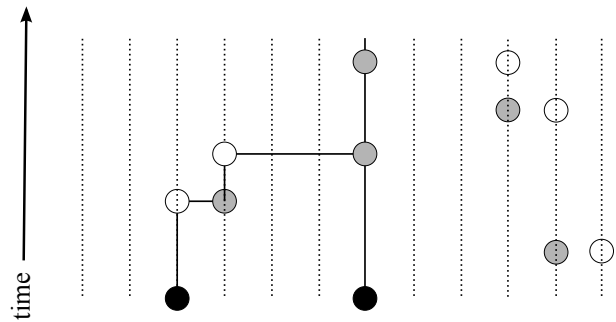,width=8cm}}
\vspace*{8pt}
\caption{\label{fig:coal-uns} Backwards-time interpretation of spatially-unstructured random copying.  Each vertical dotted line indicates one of the $M$ individuals in the population.  Shown solid are two at the present time whose ancestral lineages (solid lines) interest us. As we look backwards in time (up the page), the individual that was copied is shown shaded, and its offspring as an open circle.  At the third copying event (looking backward in time), one of the two ancestors of interest was the direct offspring of the other. At this point in time, the two lineages merge.}
\end{figure}

We may now ask for the probability that one of the two individuals was created by copying the other one.  This is $\frac{2}{M}$, because the probability that one \emph{particular} individual is the parent of another is $1/M$, and there are two ways of assigning the roles of parent and offspring to the pair of individuals.  Before such a copying event, the ancestors of the original pair of individuals are distinct; after this event the pair has a single \emph{common ancestor}.  Equivalently, the pair of ancestral \emph{lineages} coalesce at a rate $\frac{2}{M}$. See Fig.~\ref{fig:coal-uns}.  (Ref.~\cite{wak08} provides an excellent introduction to this `backwards-time' way of thinking in evolutionary dynamics).

Let us now introduce the probability $D(t)$ that two individuals randomly chosen from some \emph{present-day} population have distinct ancestors at a time $t$ in the past.  This is equal to the probability that the Poisson coalescence process that takes place at rate $\frac{2}{M}$ has not occurred by time $t$.  This is known to be \cite{dur99}
\begin{equation}
D(t) = \exp\left(-\frac{2t}{M}\right) \;.
\end{equation}
The two individuals are of different species at the present time only if they have distinct ancestors at some time $t$ in the past, and these ancestors are themselves of different species at that time.  If this earlier time is $t=0$, and the probability that a random pair of individuals are distinct at this time is $H(0)$, we have at the present time $t$
\begin{equation}
\label{Hun}
H(t) = D(t) H(0) = H(0) \exp\left(-\frac{2t}{M}\right) \;.
\end{equation}
In words, the probability that two individuals are of a different species decreases exponentially from its initial value at a rate $\frac{2t}{M}$.  As we will see below, it is the spatial analog of this result that reveals the facets of the random-copying dynamics that may or may not be affected by the presence of spatial structure.

\section{A Spatially-Structured Moran Model}
\label{stsec}

\subsection{Model definition}

\begin{figure}[th]
\centerline{\psfig{file=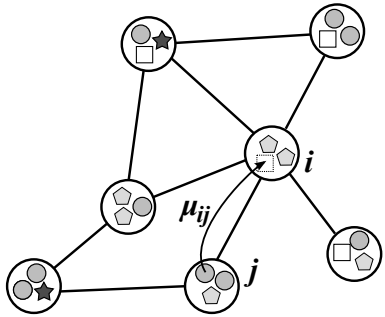,width=5cm}}
\vspace*{8pt}
\caption{\label{fig:randomcopy} Spatially-structured random copying dynamics.  At a rate $\mu_{ij}$ an individual is chosen at random from site $j$ and leaves and offspring on site $i$.  An existing individual on the receiving site (here, an member of the `square' species) is deleted to make way for the new copy.}
\end{figure}

In the previous section we described how random-copying proceeds within an unstructured population of $M$ individuals.  To obtain a spatially-structured extension, we place such a population on each site of a network of $N$ sites.  Each individual on site $j$ is copied as a Poisson process at rate $\mu_{ij}$ with the copy being placed on site $i$.  As before, a randomly-chosen individual on the receiving site is replaced so that each of the subpopulation sizes remains constant at $M$.  See Fig.~\ref{fig:randomcopy} for an illustration of these dynamics.

Note that the rate at which a copy is placed on the same site as the parent, $\mu_{ii}$, can be nonzero---in fact, we will in general require this to be the case.  Note also that the rate at which an individual is copied from site $i$ to $j$ need not be the same as the rate of copy in the opposite direction.  Indeed, one of these rates may be zero, in which case, copying between those sites is completely asymmetric.

This definition allows almost arbitrary connections between different points in space to be set up.  For example, the $\mu_{ij}$ rates could be chosen such that copying takes place between neighboring sites on a regular lattice (such as a square or triangular lattice). Alternatively, the different locations could relate to habitats that are not regularly distributed over some geographical region. Then, the magnitude of $\mu_{ij}$ would relate to how easily an offspring of a parent sampled from site $j$ could migrate to site $i$. This could depend on the distance between the two sites, but also the nature of the terrain between them, the presence or absence of waterways and so on.

Another possibility is that $i$ and $j$ relate to mobile agents. For example, in a model of language change \cite{bax06}, the frequency of a species at a given site relates to how often the user of a language uses a particular linguistic convention to signify a particular meaning.  In the language of the spatially-structured Moran model, $\mu_{ij}$ is proportional that rate at which a hearer $i$ modifies his behavior in response to an utterance produced by speaker $j$.  This rate is large when the hearer is strongly influenced by the speaker, perhaps because they interact frequently, or because the speaker has some social status that is viewed favorably by the hearer.

\subsection{Dynamics of the ancestral lineages}

\begin{figure}[th]
\centerline{\psfig{file=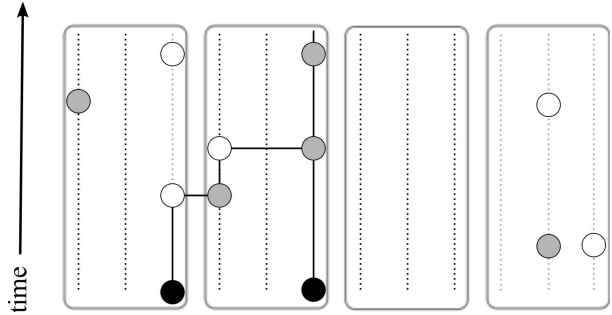,width=8cm}}
\vspace*{8pt}
\caption{\label{fig:coal-st} Backward time representation of the spatially-structured random copying dynamics.  As in Fig.~\ref{fig:coal-uns}, dotted lines represent individuals and solid lines ancestral lineages of two individuals of interest.  Again, shaded circles indicate the copied individual at some time in the past, and open circles their offspring.  Here, individuals are grouped according to their site.  When a parent is from a different site to the offspring, the ancestral lineage `migrates' to the parent's site.  In principle, lineages on different sites can coalesce; however, it is typically to take the population size on each site, $M$, large, in which case coalescence events between lineages on the same site occur with high probability (see text).}
\end{figure}

The key to understanding the effect of spatial structure on the random-copying dynamics is to identify the spatial generalization of the backward time process in which ancestral lineages coalesce.  In the case of a single unstructured population, within which each individual is copied at unit rate, we had that two lineages coalesce at a rate $\frac{2}{M}$.  Since each subpopulation $i$ is unstructured, and copying takes place within it takes place at rate $\mu_{ii}$, we now have a coalescence rate between two lineages in subpopulation $i$ at rate $\frac{2\mu_{ii}}{M}$.  However, it is also possible that, looking back into the past, an individual was created by copying from another subpopulation.  The effect of this is for an ancestral lineage to hop (or migrate) to another site on the network.  Specifically, an ancestor hops from site $i$ to site $j$ at a rate $\mu_{ij}$. The spatially-structured ancestral dynamics are illustrated in Fig.~\ref{fig:coal-st}.

To summarize, the history of two individuals sampled from the present-day population can be described in terms of a pair of coalescing random walkers. Each walker hops from site $i$ to $j$ at rate $\mu_{ij}$, and if they are on the same site $i$, they coalesce at rate $\frac{2\mu_{ii}}{M}$.  In principle, it is possible for a pair of walkers on different sites to coalesce.  However, it is customary to consider the case of large subpopulation sizes $M$.  In this regime, the coalescence rate is suppressed relative to the migration rates.  Put another way, the strength of migration relative to coalescence can be expressed in terms of the set of rates $m_{ij}$ defined through
\begin{equation}
\label{mscale}
\mu_{ij} = \frac{m_{ij}}{M} \;.
\end{equation}
These $m_{ij}$ parameters should then be compared with the coalescence rates $c_i$, that we define as
\begin{equation}
\label{cscale}
\mu_{ii} = \frac{c_i}{2}
\end{equation}
so as to dispense with annoying factors of $2$ that would otherwise appear in many expressions.  If the $m_{ij}$ are large compared to $c_i$, copying \emph{between} sites is the dominant process, whereas if they are small, copying \emph{within} sites dominates.  In practice, one tends to fix the parameters $m_{ij}$ and $c_i$, and assume that the subpopulation size $M$ is large.  Then, the rate at which coalescence between ancestors on different sites is proportional to $1/M^2$, and is sufficiently small (compared to migration and on-site coalescence processes) that these contributions to the dynamics can be neglected.  In mathematical population genetics, this coalescing random walk process is called \emph{the structured coalescent} \cite{wak08}.

The results of Ref.~\cite{bly10} are couched in terms of two timescales of the coalescing random walk process.  First, we may consider the fate of one of the two walkers.  Given that it starts on site $i$, after some time $t$ (looking into the past), it has some probability $Q_k(t)$ of being on site $k$.  After sufficiently long time, this distribution takes the form
\begin{equation}
\label{Qk}
Q_k(t) \sim Q_k + a_{ik} {\rm e}^{-t / T_1}
\end{equation}
where here $\sim$ implies increasing accuracy of the right-hand side as $t$ increases.  At very large times, $Q_k(t)$ approaches the time-independent (stationary) distribution $Q_k$.  The timescale over which this stationary state is reached is given by the parameter $T_1$.  One can calculate these quantities from the eigenvectors and eigenvalues of the matrix of hop rates between sites. We present details in the Appendix.  Loosely speaking, $T_1$ can be interpreted as the time required for a single random walker tracing the ancestry of an individual to have explored the \emph{entire} network of $N$ sites.

As in an unstructured population, the decay of diversity on a structured population can be quantified in terms of the probability that two randomly-chosen individuals are distinct.  Recall from Section~\ref{stsec} that this could be expressed in terms of the probability that two ancestral lineages have not coalesced by some time $t$.  To understand how diversity decays in a structured population, we need to consider a more complicated quantity, namely the probability that  the ancestors of two individuals currently on sites $i$ and $j$ have not coalesced by time $t$, and occupy sites $k$ and $\ell$ respectively.  At late times, this distribution decays to zero as
\begin{equation}
\label{Qkl}
Q_{k\ell | ij} \sim P_{ij} Q_{k\ell} {\rm e}^{-t/T_2} \;.
\end{equation}
Here the quantities $P_{ij}$, $Q_{k\ell}$ and $T_2$ are related to eigenvalues and eigenvectors of a matrix describing the hop and coalescence rates of a pair of random walkers on the network (see Appendix).

The key points are as follows. $Q_{k\ell}$ is the probability that the pair occupies sites $k$ and $\ell$ at late times \emph{given} that they have not coalesced.  This distribution is called the \emph{quasistationary} distribution in Ref.~\cite{bly10}.  The rate of decay of this quasistationary state is inversely proportional to the timescale $T_2$.  This time can therefore be interpreted as that required for the two random walkers to meet each other and coalesce.

\subsection{Separation of timescales}

The main result of Ref.~\cite{bly10} is that if the time required for a single walker to explore the network, $T_1$, is much shorter than that required for two walkers to coalesce, $T_2$, the frequency of a species of interest across the entire structured population is governed by the same stochastic equation of motion as a species in an unstructured population.  The only difference is that the single characteristic timescale in the unstructured population, given by the coalescence rate between two ancestral lineages, is replaced by the coalescence timescale in the quasistationary state, $T_2$.  When this occurs, a \emph{separation of timescales} is said to apply.  The essential point is that when $T_2 \gg T_1$, the coalescence time is by far the longest of \emph{all} timescales in the dynamics, and hence dominates the history of the present-day population. In the next section we provide examples of networks on which such a separation of timescales is and is not obtained.

The most straightforward interpretation of the above result is that spatial structure has no effect on the random copying dynamics when there is a separation of timescales (other than to modify the characteristic timescale).  This turns out to be nearly correct, but spatial structure can nevertheless have subtle but important residual effects.

Let us introduce the diversity measure $H_{ij}(t)$, that is the probability that two individuals chosen at random from sites $i$ and $j$  are of different species.  This can be written as
\begin{equation}
H_{ij}(t) = \sum_{k\ell} Q_{k\ell | ij} (t) H_{k\ell}(0) \sim P_{ij} {\rm e}^{-t/T_2} \sum_{k\ell} Q_{k\ell} H_{k \ell}(0)
\end{equation}
because the probability that this pair has not coalesced into a single ancestor and occupies sites $k$ and $\ell$ is $Q_{k\ell | ij}(t)$, and that the probability a pair of individuals on those sites at the beginning of time are distinct is $H_{k\ell}(0)$.  Let us suppose that the distribution of species at time $t=0$ does not exhibit any spatial correlations: that is, $H_{k\ell}(0) = H(0)$ for any pair of sites $k$ and $\ell$.  Then, because $Q_{k\ell|ij}$ is a probability distribution over pairs of sites $k$ and $\ell$, we have $\sum_{k\ell} Q_{k\ell|ij} = 1$.  Then we find that
\begin{equation}
\label{Hst}
H_{ij}(t) \sim H(0) P_{ij} {\rm e}^{-t/T_2} \;.
\end{equation}

We can compare this result with its counterpart for the unstructured population (\ref{Hun}).  As expected, the characteristic timescale of coalescence in the unstructured population $\frac{M}{2}$, has been replaced by the corresponding quantity for the structured population, $T_2$.  However, we see the appearance of a new factor $P_{ij}$ that depends on the spatial location of the sampled pair.

This factor is entirely due to interactions between the ancestral lineages that occur on the much shorter timescale $T_1$.  The picture here is that, looking back in time, there is a very short period over which the locations of lineages become randomized, reaching the quasistationary distribution $Q_{k\ell}$ if they do not coalesce.  However, there is some probability that during this initial \emph{scattering phase} (a term coined by Wakeley \cite{wak08}), the lineages coalesce.  This probability is proportional to $P_{ij}$.  In fact, when there is a total separation of timescales, $T_1/T_2 \to 0$, $P_{ij}$ \emph{is} the probability that two lineages avoid coalescence before entering the quasistationary state. In this quasistationary state, both the distribution of a single ancestor, and of a pair of ancestors conditioned on not having coalesced, are stationary.  From this point onwards, the network structure is averaged out due to the fast characteristic timescale of the hopping process ($T_1$) relative to the coalescence process ($T_2$).

We thus see that, despite the separation of timescales, there can nevertheless be spatial variation in diversity due to the dynamics that takes place on the shorter timescale.  When the coefficient $P_{ij}$ is close to unity, the diversity, as measured by sampling from sites $i$ and $j$, is somewhat similar to what would be observed in an unstructured population.  As we will see in the next section, this is typically the case when the sites $i$ and $j$ are far apart.  Conversely, on nearby sites, one may find that $P_{ij}$ is significantly reduced, indicating that individuals are likely to be more similar on those sites (as one might expect).  In these instances, the application of results and methods of analysis from spatially-unstructured models may be flawed.

\section{Random Copying on Example Networks}
\label{exsec}

We now illustrate the general results outlined in the previous section with some explicit examples of models with spatial structure.  These models have therefore been chosen to be simple but illustrative, as opposed to realistic.

\subsection{The fully-connected network}

\begin{figure}[th]
\centerline{\psfig{file=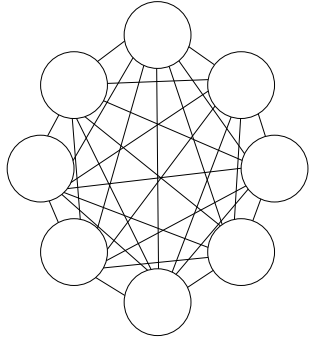,width=4cm}}
\vspace*{8pt}
\caption{\label{fig:fullcon} A fully-connected network.  Open circles indicate the sites of the network, and lines connect pairs of sites between which copying may take place.  Here, the offsping of any individual may be placed on any site (including that of its parent).}
\end{figure}

The very simplest example of a model with spatial structure is a fully-connected network, illustrated in Fig.~\ref{fig:fullcon}.  In this model, copying can take place between any pair of sites, and the only distinction that is made is whether copying takes place within the same site, or between two different sites.

More precisely, the copying rates are defined in terms of two parameters $c$ and $m$ through
\begin{equation}
\label{muful}
\mu_{ij} = \left\{ \begin{array}{ll} \frac{c}{2} & i = j\\
\frac{m}{(N-1)M} & i \ne j \end{array} \right. \;.
\end{equation}
Here, we have adopted the parametrization introduced in Eqs.~(\ref{mscale}) and (\ref{cscale}).  The on-site coalescence rate $c$ is thus common to all sites. The factor of $N-1$ that appears in the between-site copying rates ensures that the \emph{total} hop rate of an ancestor out of any site is independent of the size of the network.  This allows results for different network sizes to be compared more easily.

The various quantities that appear in the expressions (\ref{Qk}) and (\ref{Qkl}) can be calculated exactly for this model \cite{bly10}.  Although this calculation is reasonably straightforward, it is nevertheless a little lengthy so we omit the details here in favor of interpreting the results.

First let us compare the characteristic timescales $T_1$ and $T_2$.  The relaxation time for a single lineage, $T_1$, is given by
\begin{equation}
T_1 \sim \frac{M}{m}
\end{equation}
where here $\sim$ means ``plus a correction that vanishes as the network size is increased''.  The key thing to note here is that this relaxation time is independent of the network size $N$.  We can understand this in the following way.  After a single hop, an ancestor is equally likely to be found anywhere (apart from the site it started on).  The waiting time for this hop is set up to be independent of the network size, and so the ancestor is equally likely to be anywhere after a short time that does not depend on $N$.

Meanwhile, the characteristic timescale of the two-lineage coalescence process in the quasistationary state is
\begin{equation}
T_2 \sim \frac{2m+c}{2mc} NM \;.
\end{equation}
In contrast to $T_1$, this timescale increases linearly with the network size $N$.  The reason for this is the following.  We expect $T_2$ to be proportional to the number of hops needed for the two ancestors are on the same site, since this is a precondition for them to coalesce.  If both ancestors are more-or-less equally likely to be found on any site, then after any one ancestor hops, the probability that the target site contains the other ancestor is $1/N$.  We therefore expect order $N$ hops for the two ancestors to meet.  Thus, on large networks, we have that $T_1/T_2 \sim 1/N$, and that a separation of timescales is obtained in this limit.

We now examine the quantities $P_{ij}$ and $Q_{k\ell}$ that appear in (\ref{Qkl}) and reveal how spatial structure manifests itself.  For large $N$, one finds that \cite{bly10}
\begin{equation}
P_{ij} \sim \phi_{ij} \quad\mbox{and}\quad
Q_{k\ell} \sim \frac{1}{N^2} \phi_{k\ell}
\end{equation}
where
\begin{equation}
\phi_{ij} = \left\{ \begin{array}{ll} \frac{2m}{2m+c} & i=j \\ 1 & i\ne j \end{array} \right. \;.
\end{equation}

Using (\ref{Hst}) we immediately see that if two individuals are sampled from different sites $i$ and $j$, they have the same probability of belonging to different species as in an unstructured model, albeit defined on a different timescale.  On the other hand, a pair of individuals sampled from the same site are a factor $\frac{2m}{2m+c}$ less likely to be of different species than a pair sampled from different sites.  In this model, if we construct a sample from individuals all taken from different sites, we would expect their properties to be exactly the same as in an unstructured population.

These explicit expressions provide a little more insight into the nature of the quasistationary state and how it is reached.  The randomization (scattering) of the ancestral lineages takes a time of order $T_1$, which as we have seen is independent of $N$ for this model.  By making $N$ large, we can prolong the decay of the quasistationary state, such that at intermediate times $T_1 \ll t \ll T_2$, a pair of ancestors are very unlikely to have coalesced, unless they started on the same site, in which case they will have coalesced with a probability close to $1-P_{ii} = \frac{c}{2m+c}$.  From the theory of Poisson processes \cite{dur99}, one can determine that this is the probability that the two ancestors coalesce before one of them migrates to another site.  If one lineage does hop, they are unlikely to be on the same site again for a time of order $T_2$, and therefore from this point, the quasistationary state will be entered with high probability.

In this quasistationary state, the two ancestors are most likely to be found on different sites.  The probability of finding them on the same site, given that they have not previously coalesced, is $\sum_i Q_{ii} = \frac{1}{N} \frac{2m}{2m+c}$. We see this probability is dramatically reduced if the coalescence rate $c$ is large:  the fact that pairs of ancestors coalesce rapidly when on the same site means one is unlikely to see such pairs in the quasistationary state, as one might expect.  Generically, one expects such ``holes'' in the quasistationary probability distribution for nearby pairs.

\subsection{Random copying on a ring}

\begin{figure}[th]
\centerline{\psfig{file=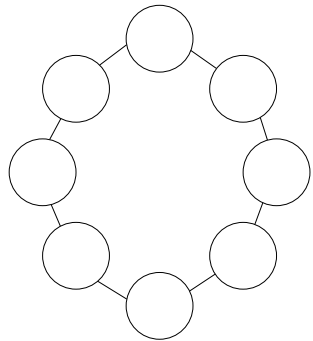,width=4cm}}
\vspace*{8pt}
\caption{\label{fig:ring} The ring network.  Here, an offspring copy may be placed only on neighboring sites, clockwise or anticlockwise, around the ring, or on the parent site.}
\end{figure}

One case where a separation of timescales is not obtained is on a one-dimensional chain of $N$ sites wrapped around to form a ring, as shown in Fig.~\ref{fig:ring}.  To understand why is reasonably straightforward.  Recall that $T_1$ is the time required for a single ancestor to explore the entire ring through a sequence of hops from a site to one of its two neighbors.  Now, $T_2$ is expected to be proportional to the time needed for a pair of ancestors to find each other.  This can be determined by examining the relative distance between the two ancestors.  This can increase by one, if one of the ancestors hops away from the other, or decrease by one, if one of them hops towards the other.  This is exactly the same hopping process as that experienced by a single ancestor's position on the ring, except that because there are two ancestors, the rate at which the relative position changes is twice that of either of the absolute positions.  Thus one expects $T_2$ to be related to $T_1$ by a constant factor that does not strongly depend on the system size.  As we will see explicitly in the next section, a separation of timescales will not be obtained, no matter how large the ring is made.

\subsection{Random copying on a small-world network}

\begin{figure}[th]
\centerline{%
\psfig{file=ring.eps,width=4cm}\hspace{1em}\psfig{file=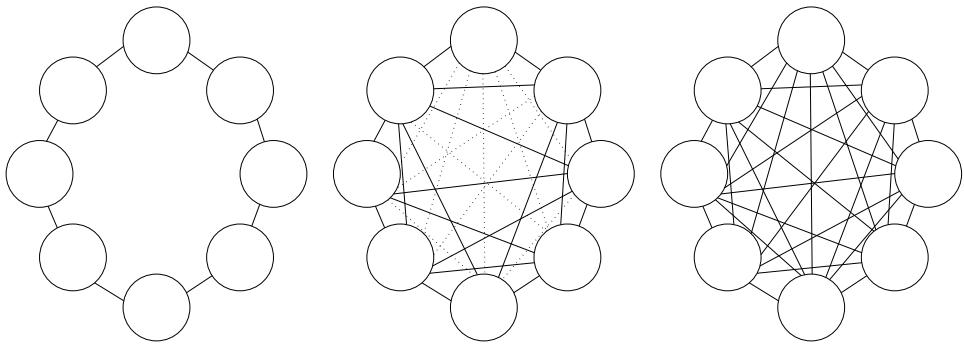,width=4cm}\hspace{1em}\psfig{file=fullcon.eps,width=4cm}
}
\vspace*{8pt}
\caption{\label{fig:sw} Small-world network (center) as an interpolation between the ring (left) and fully-connected network (right).  The small-world network is constructed by only activating a fraction $p$ of the available long-range connections.  In the center figure, unactivated links are shown dotted.}
\end{figure}

The fully-connected network and the ring are lie at opposite extremes of a continuum of network structures collectively known as \emph{small-world networks} \cite{wat98}.  Starting with a ring of $N$ sites, one can construct the fully-connected network by iteratively adding links between randomly-chosen pairs of sites that are not directly connected until such time that all possible links have been added. One way to construct a small-world network is to follow the same sequence of steps, but stop after some pre-determined number of links has been added.  See Fig.~\ref{fig:sw}. These randomly-added links we refer to as \emph{long-range} links, to distinguish them from the nearest-neighbor links that are present before they are added. (In the original work on small-world networks \cite{wat98}, the long-range links were formed by rewiring the original nearest-neighbour links as opposed to adding them. Both methods of construction are understood to lead to networks with broadly similar properties---see e.g., \cite{new99}).

In addition to the size of the network, $N$, small-world networks are further characterized by a parameter $p$ defined as the mean fraction of available sites to which any node is connected by long-range links.  If $p=0$, only the nearest-neighbor links are present, and the ring is recovered. Conversely, if $p=1$, all possible links are present, and the fully-connected network is obtained. Since we obtain a separation of timescales in the limit $p=1$, but do not when $p=0$, the behavior of the relevant timescales at intermediate values is of interest.  More generally, these intermediate cases have both local structure and a short mean distance between any pair of nodes \cite{wat98}. These characteristics are believed to be shared with social networks, for example---although we do not mean to imply that the small-world network as described here is an \emph{accurate} representation of the network of human interpersonal relationships. We will therefore also be interested in seeing how these two properties impact on the random-copying dynamics.

The copying rates for this model are defined analogously to those for the fully-connected network (\ref{muful}).  Recall that there, a factor $(N-1)$ was included in the between-site copying rate so that the total rate at which a copy is received by a site is independent of the network size.  The generalization of this idea to the case where different sites may have different numbers of neighbors is to ensure that the total rate of copying into a site depends neither on the network size, nor on the number of neighbors.  This implies copying rates of the form
\begin{equation}
\mu_{ij} = \left\{ \begin{array}{ll} \frac{c}{2} & \mbox{if $i=j$} \\
\frac{m}{Mz_i} & \mbox{if site $i$ is connected to $j$} \end{array}\right.
\end{equation}
where $z_i$ is the \emph{degree} of site $i$, i.e., the number of sites it is connected to.  We remark that this choice essentially corresponds to the \emph{voter model} which has been studied widely in the mathematics and physics literature (see e.g.~\cite{cas09} which contains a comprehensive review of the voter model in the context of social dynamics, and \cite{bly10} for a precise statement of how to obtain voter model dynamics within the more general model described here).  A property of these rates is that the rate of copying from a poorly-connected site onto a well-connected site is less than the other way round.  As a consequence, well-connected sites tend to have a bigger effect on the overall dynamics of the random-copying process than poorly-connected sites \cite{cas09}.

We are not aware of any methods that allow the characteristic timescales $T_1$ and $T_2$, or the spatially-dependent quantities $P_{ij}$ and $Q_{k\ell}$, to be calculated exactly.  As we noted previously, these are related to eigenvalues and eigenvectors of a pair of matrices whose forms are given in the Appendix.  We thus resort to numerical methods for obtaining these, details of which are also provided in the Appendix.

\begin{figure}[th]
\centerline{\psfig{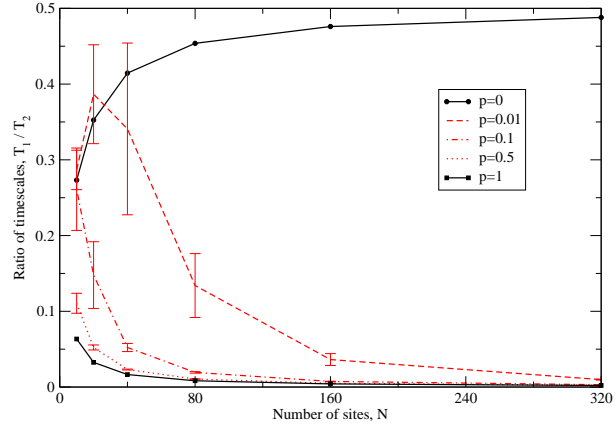}}
\vspace*{8pt}
\caption{\label{fig:ratio} Ratio of characteristic timescales $T_1/T_2$ as a function of network size $N$ at various $p$ for $m=c=1$.  The case $p=0$ (uppermost solid curve) corresponds to the ring, whilst $p=1$ (lowermost solid curve) corresponds to the fully-connected network.  For all nonzero $p$, the ratio of timescales eventually vanishes as $N$ is increased, indicating a separation of single-ancestor relaxation time and the lifetime of the quasistationary two-ancestor state.}
\end{figure}

We first determine the conditions under which a separation of timescales is arrived at on a small-world network.  In Figure~\ref{fig:ratio} we plot the ratio $T_1/T_2$ as a function of $N$ for different values of $p$, and with $m=c=1$ in all cases\footnote{The choice $m=c=1$ puts both the coalescence and migration processes on exactly the same timescale, and it is in this regime that one expects these processes to interact in the most nontrivial way.}.  Above we showed that on the fully-connected network, $p=1$, we have $T_1/T_2 \sim 1/N$ for large $N$.  That is, the separation of timescales is found on sufficiently large networks.  On the other hand, we argued that this ratio remains finite even for large networks when $p=0$.  The solid lines in Figure~\ref{fig:ratio} correspond to these extreme cases, and the predicted behavior is indeed observed within the numerical calculations.

For values of $p$ larger than about $0.1$ (that is, when any site is directly connected to one in ten of the other sites), we find the ratio $T_1/T_2$ decays to zero in much the same way as it does for the fully-connected network.  The case of small $p$ is most interesting.  As the size of the network is increased, the ratio $T_1/T_2$ initially \emph{increases}, just as it does for random copying on the ring ($p=0$). Then, one the system size is sufficiently large, the ratio decays towards zero, as it does for larger $p$.

We will exploit known properties of small-world networks \cite{new99,men00} to understand the crossover from ring-like behavior to that of the fully-connected network at some intermediate $N$ when $p$ is small.  Recall that in the construction of the small-world network, each site is connected to one of the $N-2$ initially non-neighbor sites with probability $p$.  The probability that a site has no long-range links is that $(1-p)^{N-2}$.  The typical distance between two sites with at least one long-range link is then $1/(1-[1-p]^{N-2})$.  The peak in the ratio of timescales occurs at $N\approx 20$ for $p=0.01$, suggesting that when the typical distance between sites with long-range links is less than about $5$, the density of such connections is sufficiently high that it is effectively equivalent to fully-connected network.  More generally, a large network is known to exhibit a \emph{small-world transition} \cite{new99,men00} between the ring-like and fully-connected behaviors at a value of $p \propto 1/N$.  As $N$ is increased, the value of $p$ needed for the long-range links to be so sparse that the ring-like behavior is seen decreases towards zero.  That is, if one has any non-zero $p$, then the network can be made sufficiently large that the number of long-range links allows it to be explored rapidly, and much more rapidly than it takes for two random walkers to find each other and interact.

\begin{figure}[th]
\centerline{\psfig{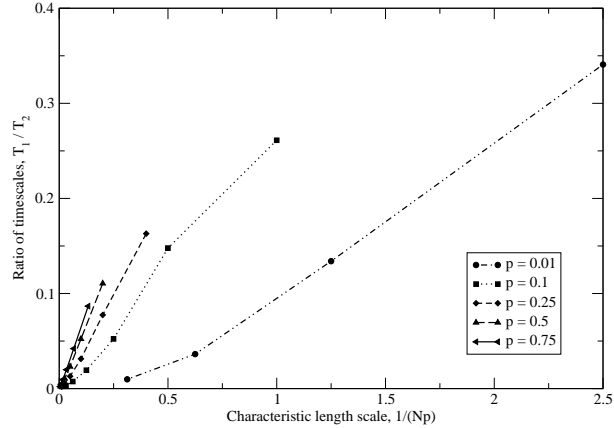}}
\vspace*{8pt}
\caption{\label{fig:ratioell} Ratio of characteristic timescales $T_1/T_2$ as a function of the length $\ell = Np$ that characterizes the typical distance between sites with long-range connections. Each line shows a different value of $p$; $m=c=1$ as previously.  We see that the timescale ratio vanishes as $\ell$ is decreased, but that we do not find that the timescale ratio is a universal function of $\ell$.}
\end{figure}

These considerations suggest it is worth plotting the timescale ratio $T_1/T_2$ as a function of the distance $\ell$ between long-range connections, which is approximately given by $Np$.  These data are shown in Figure~\ref{fig:ratioell}. What is clear is that the ratio of timescales vanishes---in some cases rapidly---as $\ell$ decreases.  We do not, however, find that $T_1/T_2$ depends only on $N$ and $p$ through $\ell$: the curves for different $p$ do not sit on top each other.  However, this does not alter the fact that anything that serves to reduce the typical distance between long-range connections (either increasing $N$ or $p$) leads to a dynamics with a more mean-field character, even though only a small fraction of all possible links may be present.

We now turn to the structure factors $P_{ij}$.  Recall from the discussion around Eq.~(\ref{Hst}) that $P_{ij}$ tells us how likely a pair of individuals on sites $i$ and $j$ are to have a common ancestor from the recent past, relative to a randomly pair of individuals drawn from an unstructured population.  Specifically, if $P_{ij}<1$, the two individuals are more likely to have a recent common ancestor---and therefore of the same species---than a pair sampled from an unstructured population.  On the other hand if $P_{ij} \approx 1$, spatial structure does not affect the expected diversity within a sample.

\begin{figure}[th]
\centerline{\psfig{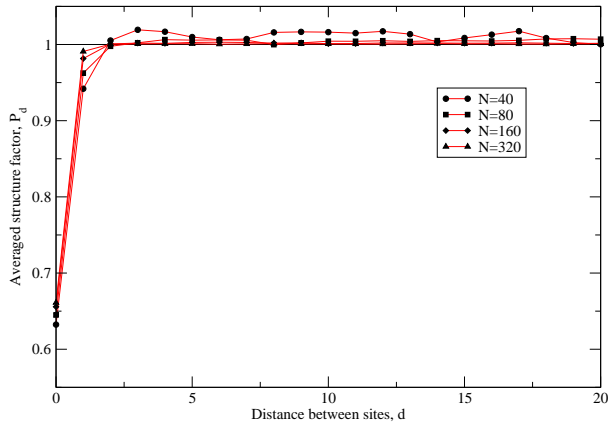}}
\vspace*{8pt}
\caption{\label{fig:structures0.1} Structure factor $P_d$ as a function of the distance $d$ between pairs of sites around the edge of the network for $p=0.1$, $m=c=1$.  When $P_d=1$, the expected diversity is the same as in an unstructured population.  At all the system sizes displaced, $P_d$ differs only from $1$ when individuals are only a few sites apart.}
\end{figure}

We examine first the case of $p=0.1$ and $m=c=1$, where we see from Fig.~\ref{fig:ratio} that on networks of $N=40$ or more sites, the coalescence timescale $T_2$ is at least twenty times longer than the relaxation time $T_1$.  On single realizations of small-world networks of different sizes, we define the average structure factor $P_d$ for two sites a distance $d$ apart (as measured on the original ring structure before the long-range links are added) as
\begin{equation}
P_d = \frac{1}{N} \sum_{i=1}^{N} P_{i,i+d} \;.
\end{equation}
Note we are using periodic boundary conditions, so that $P_{i,j} = P_{i+N,j} = P_{i,j+N}$ for any $i$ and $j$, and that valid values of the distance $d$ are the integers $d=0,1,2,\ldots$ less than or equal to $N/2$.  Then, the quantity $P_d$ tells us the likelihood for two individuals to find their common ancestor at some time after the initial relaxation time $T_1$, given that they are currently a distance $d$ apart.  We plot $P_d$ as a function of $d$ at different $N$ in Fig.~\ref{fig:structures0.1}. We see that for sufficiently large $d$, $P_d \approx 1$, indicating that two individuals sampled at least three apart from each other have the same diversity statistics as in an unstructured population.  On the other hand, when two individuals are sampled from the same sites, or from two neighboring sites, there is a much larger probability that these individuals have a common ancestor from the recent past (i.e., on the timescale $T_1$).

\begin{figure}[th]
\centerline{\psfig{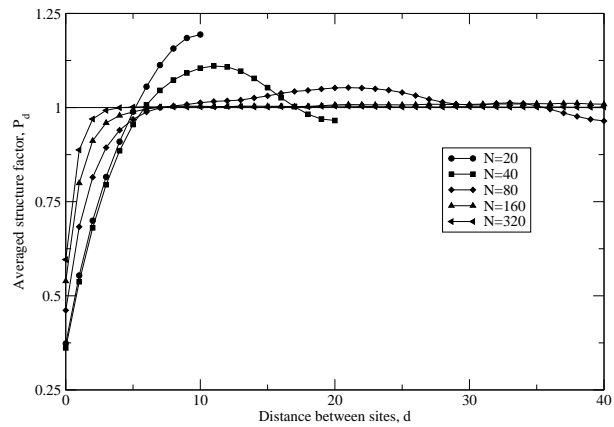}}
\vspace*{8pt}
\caption{\label{fig:structures0.01} Structure factor $P_d$ as a function of the distance $d$ between pairs of sites around the edge of the network for $p=0.01$, $m=c=1$.  Here there are significant deviations from the unstructured baseline ($P_d=1$) on small networks ($N<80$). This is the regime in which the characteristic timescales are not well-separated.}
\end{figure}

We now contrast the case of $p=0.01$ and $m=c=1$, where we inferred from Fig.~\ref{fig:ratio} a ring-link behavior on networks of approximately 40 sites or fewer.  It is on these small networks that we see significant deviation of $P_d$ from unity in Fig.~\ref{fig:structures0.01}.  Since one does not have a separation of timescales on these networks, $P_d$ cannot be straightforwardly interpreted as a probability (as evidenced by the fact that it can greatly exceed unity).  Here we must instead interpret $P_d$ as a \emph{relative} probability---that is, the probability that the two ancestral lineages have not coalesced after a time of order $T_2$, relative to two chosen from an unstructured population.  Since the \emph{absolute} probabilities of not having coalesced after such long times may be quite small, their ratios may exceed unity.  On these small networks, we cannot simply apply results from unstructured populations, because we do not have the required separation of timescales.  Even if there were a separation of timescales, the expected amount of diversity in a sample would be strongly dependent on the distances between the sites from which samples were drawn.  Meanwhile, on the larger networks, we see that $P_d$ assumes a profile similar to that seen for larger $p$. That is, if individuals are taken from locations closer together than about 5 sites, they are more likely to be of the same species than if they are sampled from further apart (or from an unstructured population).

In summary, the results of this section suggest that if the density of long-range links exceeds a value that decreases with the size of the network, one is likely to find the separation of timescales that implies that a sample's history is dominated by a long-lived quasistationary state where the location of ancestors is randomized, and coalescence of lineages takes place a constant rate.  Furthermore, it would appear that the expected amount of diversity seen within a sample of individuals will be the same as that seen in an unstructured population (and in which lineages coalesce at the same rate), as long as each individual in the sample is initially far enough away from the others that they are unlikely to have coalesced in the recent past.  We have illustrated the latter point explicitly with samples of size $2$ here.  It would be interesting to see whether the statement also holds for larger samples, and if a small density of long-range links is sufficient for the separation of timescales to emerge on a more general class of networks than small-world networks.

\section{Discussion and Conclusion}
\label{concs}

In this work we have set out to investigate the effects of spatial structure on evolution by random copying, which is typically assumed to operate in a non-spatial setting.  We aimed in part to answer two questions.  First, whether a fit of an unstructured random-copying model to empirical data implied that all the (potentially unreasonable) assumptions of such a model must necessary hold; and second, whether a bad fit can be ascribed to fitness differences between the species.  As advertized in the introduction, our study of random copying in space allows one to answer both questions negatively.

As we have seen, when there is a separation of timescales there exists a long-lived quasistationary state during which a spatially-structured model behaves exactly as its unstructured counterpart.  We illustrated this through the behavior of a pair of lineages.  In the unstructured model, these lineages do not move in space, and coalesce as a Poisson process with a constant rate.  In the structured model, and under the separation of timescales, the lineages hop between sites on a rapid timescale.  This, in effect, performs a spatial averaging that leads to coalescence occurring as a Poisson process with a constant rate, but one that may differ from that of an unstructured population of the same size.  Any properties that depend only on the dynamics within the quasistationary state would then be identical to those of an unstructured model.  As an example, we cited a measure of diversity obtained by sampling from sites sufficiently far from one another that copying events from the recent history do not affect the expected amount of diversity.  Many other properties of the unstructured model are available from classical work in mathematical population genetics \cite{cro70}.

However, we have also seen two mechanisms by which spatial structure can lead to behavior that is different from that predicted by an unstructured model.  The first is when recent coalescence events contribute to the history of a sample, in addition to those from the quasistationary state.  This leads to a lower level of diversity in a sample than one would expect under a spatially-unstructured model, and can occur even under a separation of timescales when samples are taken from nearby sites.  The second is when there is no separation of timescales, under which circumstances the history is not dominated by a single, long-lived quasistationary state.  Then, one would not expect the spatially-unstructured model to act as a proxy for a spatially-structured model.  Thus departures from the predictions of the unstructured random-copying process do not necessarily imply that there are fitness differences between species, since in the spatially-structured random-copying process, all species are treated equally.

There is an argument that spatial structure of the type we have described here introduces a form of selection, in that individuals that find themselves on a site that is copied from frequently are `fitter' than those on a site that is infrequently copied from.  However, despite differences in reproduction rates, this does not count as selection in the standard sense because the ability to reproduce more rapidly is not inherited by offspring from their parents (Hull \cite{hul88}, for example, gives a careful definition of selection).  Nevertheless, there are interpretations of the random copying dynamics, for example in the context of language change, where it is conceptually fruitful to think of this variation in total copying rates between sites as a form of selection that is distinct from the classical fitness of a species \cite{bax09}.

In this work we have restricted ourselves to the case where no new species may enter the population as it evolves.  We thus conclude with a few remarks on innovation.  Typically, innovation is incorporated into the random-copying process by there being some probability of replacing an existing individual with one of a completely new species, or one that is taken from a fixed external (`mainland') population \cite{hub01}.  In the backward-time picture of ancestral lineages, this amounts to a rate at which mutations can occur along branches \cite{wak08}.  In an unstructured model, this mutation rate is typically assumed to be a constant.  In a spatially-structured version, one may reasonably allow the mutation rate to vary with space.

Based on the discussion in this work, we can conjecture three regimes according to the rate of mutation (assuming that the separation of timescales discussed in this work holds).  If mutation occurs on the same timescale as the relaxation to the quasistationary state, then coalescence events in the quasistationary state will have no effect on the diversity seen in the present-day population. This is because at least one mutation event will have occurred with high probability since the time of such a coalescence event.  At the other extreme, mutation occurs at a rate that is much slower than the quasistationary coalescence rate.  In this case, mutation can be ignored, because the probability of any mutation occurring in the time since the most recent common ancestor of a present-day population is found is very small.  In the intermediate case, mutation occurs on the same timescale as the quasistationary coalesence events, which in turn is much slower than the process by which a lineage explores the entire network.  In this case, we anticipate that a spatially-varying mutation rate could then be replaced with a spatial average weighted by the stationary distribution for a single lineage.  This would allow direct application of results from spatially-unstructured models with mutation to the spatially structured case. This would include, for example, Ewens' sampling formula that forms the basis of statistical tests for selection \cite{ewe04,alo06}.  The application of such tests---for example, to the example of baby names used to illustrate the random copying dynamics at the start of this paper---could provide one means to obtain a better understanding of the interplay between selection, mutation and drift in a cultural evolutionary context.

\appendix

\section{Matrix equations for characteristic timescales and structure factors}

In this Appendix we explain how to set up the matrix equations from which the
characteristic timescales $T_1$ and $T_2$, along with the structure factors
$P_{ij}$ and $Q_{k\ell}$, appearing in Eqs.~(\ref{Qk}) and (\ref{Qkl}) are determined.

The starting point is the set of copying rates $\mu_{ij}$ that collectively define
a spatial structure and random copying dynamics upon it. We will take these rates to be expressed
in terms of the parameters $m_{ij}$ and $c_i$ that appear in Eqs.~(\ref{mscale}) and (\ref{cscale})
respectively.

We examine first the distribution of a single ancestor $Q_k(t)$, which asymptotically has
the expression (\ref{Qk}).  Our task is to construct the master equation for this distribution.
This is achieved by noting that, given the distribution at some time $t$, the probability $Q_k(t)$ increases at rate $Q_i(t) \mu_{ik}$ as a result of a lineage hopping onto site $k$ from site $i$, and decreases at rate $Q_k(t) \mu_{ki}$ through hops in the opposite direction.  The master equation is then obtained by summing over all possible $i$:
\begin{equation}
\frac{{\rm d} Q_k(t)}{{\rm d} t} = \sum_{i} Q_i(t) \mu_{ik} - \sum_{i} Q_k(t) \mu_{ki} \;.
\end{equation}
This can be written as a matrix equation
\begin{equation}
\frac{{\rm d} Q_k(t)}{{\rm d} t} = \frac{1}{M} \sum_{i} Q_i(t) A_{ik} 
\end{equation}
where $A_{ij} = m_{ij}$ if $i\ne j$, and $A_{ii} = -\sum_j m_{ij}$.  Note here we have used the
relation (\ref{mscale}) between $\mu_{ij}$ and $m_{ij}$.

The stationary distribution $Q_k$ is formed by the left eigenvector of the matrix $A$ with eigenvalue zero.  Since eigenvectors are defined only up to a normalization, we must scale this eigenvector so that $\sum_k Q_k = 1$ for it to be interpretable as a probability distribution.  We assume that the set of copying rates is such that the stationary state is unique.  A sufficient condition for this is that it is possible for each individual in the population to have, at some later time, a descendant on any site of the network.  The uniqueness of the stationary state then implies that all other eigenvalues of the matrix $A$ have negative real part.  If $\lambda_1$ is the eigenvalue with largest nonzero real part, the characteristic timescale
\begin{equation}
T_1 = - \frac{M}{\Re \lambda_1} \;.
\end{equation}

Similar considerations lead to a master equation for the probability $Q_{k\ell}(t)$ that a pair of ancestors occupy sites $k$ and $\ell$ at time $t$.  (We suppress the explicit dependence on the initial condition that is present in the main text, since this is not relevant to the determination of eigenvalues and eigenvectors).  We recall that we disregard any processes that are of order $1/M^2$ or smaller following the parametrization (\ref{mscale}) and (\ref{cscale}).  This means that $Q_{k\ell}$ can increase either by a lineage hopping from some site $i$ onto $k$, or from site $j$ onto $\ell$.  It may decrease by hops in the opposite directions.  It may also decrease at a rate $c_k$ if $k=\ell$ through coalescence of the lineages.  This leads to a matrix equation of the form
\begin{equation}
\label{me1}
\frac{{\rm d} Q_{k\ell}(t)}{{\rm d} t} = \frac{1}{M} \sum_{i} Q_{ij}(t) B_{ij;k\ell} 
\end{equation}
where the elements of $B$ are
\begin{equation}
B_{ij;k\ell} = \left\{
\begin{array}{ll}
-2\sum_{n} m_{in} - c_i & \mbox{if $i=j=k=\ell$}\\
-\sum_{n} (m_{in} + m_{kn}) & \mbox{if $i=j$, $k =\ell$ but $i\ne k$}\\
m_{ik} - \sum_{n} m_{jn} & \mbox{if $i\ne k, j=\ell$} \\
m_{j\ell} - \sum_{n} m_{in} & \mbox{if $i=k, j\ne\ell$} \\
0 & \mbox{otherwise}
\end{array} \right.\;.
\end{equation}
One should not be put off by the appearance of four indices on the the matrix $B$: it can still be represented as a standard matrix with two indices $n,m$, each ranging from $1$ to $N^2$, if one takes $n=N(i-1) + j$ and $m=N(k-1) + \ell$, for example.

A property of this system of equations is that $Q_{k\ell}(t) \to 0$ as $t\to\infty$.  This is a reflection of the fact that, eventually, the two lineages will meet and coalesce (as long as each site can be reached from any other, which was assumed above for uniqueness of the single-ancestor steady state).  An equivalent statement is that all eigenvalues of $B$ have negative real part.  The eigenvalue with largest real part, $\lambda_2$, is real, and its reciprocal defines the relaxation time for the quasistationary state via
\begin{equation}
T_2 = - \frac{M}{\lambda_2} \;.
\end{equation}
The structure factors $Q_{k\ell}$ and $P_{ij}$ (also real) that appear in (\ref{Qkl}) are proportional to the corresponding left and right eigenvectors of $B$ respectively.  The normalization of these vectors is a little subtle, so we expand on this in more detail.

Let $\phi_{k\ell}$ and $\psi_{ij}$ be the unnormalized left and right eigenvectors of $B$ corresponding to $\lambda_2$, that is, any solution to
\begin{equation}
\phi_{k\ell} = \sum_{ij} \phi_{ij} B_{ij;k\ell} \quad\mbox{and}\quad
\psi_{ij} = \sum_{ij} B_{ij;k\ell} \phi_{k\ell} \;.
\end{equation}
We want to interpret $Q_{k\ell}$ as the probability distribution for a pair to be on sites $k$ and $\ell$ in the quasistationary state, so its elements must sum to unity.  Thus
\begin{equation}
\label{Qnorm}
Q_{k\ell} = \frac{1}{\sum_{nm} \phi_{nm}} \phi_{k\ell} \;.
\end{equation}
Meanwhile, for the amplitude of the decay in (\ref{Qkl}) to be correct, we must also have $\sum_{ij} P_{ij} Q_{ij} = 1$.  Thus
\begin{equation}
\label{Pnorm}
P_{ij} = \frac{\sum_{nm} \phi_{nm}}{\sum_{nm} \phi_{nm} \psi_{nm}} \psi_{ij} \;.
\end{equation}

In the main text, it was suggested that if one has the separation of timescales $T_1 \ll T_2$, the quasistationary state is entered from an arbitrary initial condition for two ancestors.  This statement is actually known to be true only if the hop rates $\mu_{ij}$ (or equivalently $m_{ij}$) satisfy a property known as \emph{detailed balance} \cite{kel79}.  The statement of detailed balance is that
\begin{equation}
Q_k \mu_{ki} = Q_i \mu_{ik}
\end{equation}
for all $i$ and $k$, in which $Q_k$ is the stationary distribution for a single ancestor.  It turns out that all the examples discussed in the main text, as well as many others of interest \cite{bly10}, satisfy detailed balance.  For a given set of rates one can test whether detailed balance is satisfied without knowing the stationary probabilities $Q_k$ exactly by applying a Kolmogorov criterion---see \cite{kel79} for details.

When detailed balance does not hold, we suspect that if $T_1 \ll T_2$, the two-ancestor distribution still relaxes to quasistationarity on the same timescale that the one-ancestor distribution relaxes.  However, in such cases, one would need to test this explicitly by finding the eigenvalue of $B$ with second-largest real part, $\lambda_3$, and checking that $\lambda_2 / \Re \lambda_3 \ll 1$.  See \cite{bly10} for further discussion on the relationship between the three timescales that one needs, in principle, to consider.

There are various standard routines for computing eigenvalues and eigenvectors of a matrix.  Life is most straightforward if a set of rates does satisfy detailed balance.  Then, one can typically calculate the stationary distribution $Q_i$ exactly without recourse to a numerical solution.  Then, one can use, for example, routines from a linear algebra library like LAPACK \cite{lapack}, or those included with the open-source GNU Scientific Library (GSL) \cite{gsl}, to find the eigenvalues of the matrix $A$.  Likewise one can use such routines to find the largest eigenvalue and corresponding eigenvectors of $B$.  Careful reading of the documentation accompanying such packages is essential for their correct operation.

One practical problem with such an approach is that the matrix $B$ is of dimension $N^2$ and can rapidly grow too large for these numerical routines to complete in a reasonable time.  Since we only require the largest eigenvalue of $B$, one can turn instead to a simple algorithm known as \emph{power iteration} \cite{gol83}.  We will describe the case where the hop rates satisfy detailed balance, since then the matrix $B$ exhibits the symmetry
\begin{equation}
\label{Bdb}
Q_i Q_j B_{ij;k\ell} = Q_k Q_{\ell} B_{k\ell;ij}
\end{equation}
which in turns implies that if one has found a right eigenvector $\psi_{ij}$ of $B$, the corresponding left eigenvector is $\phi_{ij} = Q_i Q_j \psi_{ij}$.  We begin with an initial guess for the largest eigenvector of $\psi^{(0)}_{ij}=1$ for all $i,j$.  We can iteratively improve on this estimate by repeating the following set of steps for $n=1,2,3,\ldots$:
\begin{enumerate}
\item Construct the vector \[
v_{ij} = \sum_{k\ell} M_{ij;k\ell} \psi^{(n-1)}_{k\ell} + \Delta \psi^{(n-1)}_{ij}\]
where $\Delta$ is some constant chosen such that $\Delta \ge 4 \max_i \{-m_{ii}\} + 2 \max_i \{ c_i\}$.
\item Obtain an estimate of the largest eigenvalue
\[ \lambda^{(n)} = \sum_{ij} Q_i Q_j \psi^{(0)}_{ij} v_{ij} - \Delta \;.\]
\item Obtain an estimate of the corresponding right eigenvector
\[ \psi^{(n)}_{ij} = \frac{1}{\sqrt{\sum_{nm} Q_n Q_m v_{nm}^2}} v_{ij} \;.\]
\end{enumerate}

This is essentially the method described in \cite{gol83}. It is necessary to introduce a shift of $\Delta$ on all the eigenvalues to ensure that the largest (negative) eigenvalue of $B$ has the largest \emph{magnitude} of all eigenvalues.  For definiteness, we have also written out the scalar product between the left and right eigenvectors explicitly through the weight function $Q_i$. Once the algorithm converges, one can construct $P_{ij}$ and $Q_{ij}$ by setting $\phi_{ij} = Q_i Q_j \psi_{ij}$, and using (\ref{Qnorm}) and (\ref{Pnorm}).

\end{document}